\documentclass[10pt, conference, compsocconf]{IEEEtran}

\ifCLASSINFOpdf

\else

\fi

  \usepackage[nocompress]{cite}
\usepackage{algorithm}
\usepackage{multirow}
\usepackage{graphicx}
\usepackage{algorithm}
\usepackage{algorithmic}
\usepackage{array}
\usepackage{subfigure}
\usepackage{amsthm}

\usepackage[justification=centering]{caption}
\usepackage{url}
\usepackage{cite}
\usepackage{setspace}
\usepackage{amsmath,amssymb,amsfonts}
\usepackage{algorithmic}
\usepackage{graphicx}
\usepackage{textcomp}
\usepackage{enumerate}
\usepackage{amsfonts}
\usepackage{amsmath} 
\usepackage{amssymb}
\usepackage{verbatim}
\usepackage{color}
\usepackage{float}
\usepackage{setspace}
\usepackage{threeparttable}

\usepackage{cite}

\begin{document}

\title{A Low-latency Communication Design for Brain Simulations\\ 
}

\author{Xin Du\IEEEauthorrefmark{2}, 
	Yuhao Liu\IEEEauthorrefmark{2}, 
	Zihui Lu\IEEEauthorrefmark{2}\IEEEauthorrefmark{1},~
	Qiang Duan\IEEEauthorrefmark{3},
	Jianfeng Feng\IEEEauthorrefmark{4},
	Jie Wu\IEEEauthorrefmark{2}, 
	Boyu Chen\IEEEauthorrefmark{4} and 
	Qibao Zheng\IEEEauthorrefmark{4}
	\\
	\IEEEauthorblockA{\IEEEauthorrefmark{2}\textit{School of Computer Science,}
		\textit{Fudan University,}
		Shanghai, China 
	}
	
	\IEEEauthorblockA{\IEEEauthorrefmark{3}\textit{Information Sciences and Technology Department,
		}
		\textit{Pennsylvania State University,}
		Abington PA, United States
	}
	\IEEEauthorblockA{\IEEEauthorrefmark{4}\textit{Institute of Science and Technology for Brain-Inspiblue Intelligence,
		}
		\textit{Fudan University,}
		Shanghai, China \\
		Corresponding author: Zhihui Lu(lzh@fudan.edu.cn)
	}
}

\maketitle
\begin{abstract}
Brain simulation, as one of the latest advances in artificial intelligence, facilitates better understanding about how information is represented and processed in the brain. The extreme complexity of human brain makes brain simulations only feasible upon high-performance computing platforms. Supercomputers with a large number of interconnected graphical processing units (GPUs) are currently employed for supporting brain simulations. Therefore, high-throughput low-latency inter-GPU communications in supercomputers play a crucial role in meeting the performance requirements of brain simulation as a highly time-sensitive application.
In this paper, we first provide an overview of the current parallelizing technologies for brain simulations using multi-GPU architectures. Then, we analyze the challenges to communications for brain simulation and summarize guidelines for communication design to address such challenges. Furthermore, we propose a partitioning algorithm and a two-level routing method to achieve efficient low-latency communications in multi-GPU architecture for brain simulation. We report experiment results obtained on a supercomputer with 2,000 GPUs for simulating a brain model with 10 billion neurons to show that our approach can significantly improve communication performance. We also discuss open issues and identify some research directions for low-latency communication design for brain simulations.
\end{abstract}

\section{Introduction}

Motivated by the grand challenge of understanding how information is represented and processed in brains, brain simulations that create and run computational models of a brain have become an important inter-disciplinary research topic~\cite{6750072}. Since human brain represents the peak of biological intelligence, brain simulations not only reflect the most advanced artificial intelligence but also inspire new developments in artificial intelligence technologies. 

A human brain comprises about 86 billion neurons interconnected via around 86 $\times$ $10^{3}$ billion synapses. Simulation of the inter-neuron information exchange demands a total throughput over one terabit per second. Low-latency communication is particularly crucial to brain simulations. Human cerebral cortex takes fraction of a second to complete a sequence of tasks for information exchange and process~\cite{laughlin2003communication} thus requiring millisecond-level latency for inter-neuron communications in brain simulations. 
Therefore, brain simulations are only possible upon high-performance computing platforms, which typically leverage parallel processing technologies and graphical processing units (GPUs)~\cite{van2020exploring}.


Figure~\ref{system-structure} illustrates the structure for a brain simulation system. The simulation model is a large-scale spiking neural network whose structure is inspired by the connectivity characteristics of the human brain. The model runs on a supercomputer that consists of a group of computer nodes interconnected through a network of Infiniband switches. Each computer node comprises multiple GPUs and a CPU comprising multiple NUMA (Non-Uniform Memory Access) nodes, each utilizes the NUMA architecture to better match with the multiple GPUs. Simulation is setup through two stages -- first generate a simulation model (e.g., a spiking neural network) and then disperse the model across the GPUs in computer nodes. The simulation process comprises two main aspects -- computation on GPUs for data processing and communications among GPUs for data exchange.

In order to support large-scale brain simulations with increasing number of neurons and complex models, state-of-the-art simulators leverage the multiple GPUs available in a supercomputer for performing neuron data processing in parallel to accelerate computing speed~\cite{xu2019meurep,knight2021larger}. However, when the number of simulated neurons increases to a super-large-scale, information exchange among neurons generates a huge volume of data traffic that must be transmitted between GPUs with low latency~\cite{merolla2014million,9284088}. Therefore, communications among GPUs in a supercomputer play a crucial role in a brain simulation system and low-latency communication design becomes a significant research topic for meeting the performance requirements of brain simulations.

Various research efforts have been made for reducing communication latency in brain simulations, including topology design, cross-layer optimization, and parallel computing technologies~\cite{roy2019towards}. However, experiment results have shown that the communication bottleneck issue in super-large scale simulations with more than 4 billion neurons have not been solved by the current methods. In this article, we attempt to address the problem of low-latency communications in brain simulations from a design perspective with focus on partitioning neurons across GPUs and routing traffic among GPUs. To our best knowledge, we are the first team to study the low-latency communication problem on a scale level of 10 billion neurons for human brain simulation.    

The rest of the article is organized as follows. We first briefly review the current brain simulators and related parallel technologies in Section \ref{Review}. Then in Section \ref{Guidelines}, we analyze the characteristics of communications for brain simulations and provide a list of guidelines for low-latency communication design in such a special environment. Following the guidelines we propose a partitioning algorithm for mapping neurons to GPUs and a two-level routing scheme for inter-GPU communications in Section \ref{Design}. Experiment results are presented in Section \ref{Experiment} to show the proposed partition and routing methods may improve delay performance of large-scale brain simulations. We also discuss several challenges and identify some research directions for low-latency communication design in Section \ref{Directions} before we draw conclusions at the end of the paper.

\section{Parallel Technologies for Brain Simulations} \label{Review}

\begin{table*}[!t]
	\begin{threeparttable}
		\centering
		\fontsize{10}{16}\selectfont
		\caption{Main features of available brain simulators}
		\label{tab:performance_comparison}
		\begin{tabular}{p{2.5cm}cc|cc|ccccc|ccc}
			
			\hline
			
			\multirow{1}{*}{\bf{Simulator}}&
			\multicolumn{2}{c}{\bf{mGPU}}&
			\multicolumn{2}{c}{\bf{Status}}&
			\multicolumn{5}{c}{\bf{Programming paradigm}}&
			\multicolumn{2}{c}{\bf{SNN Models}}\cr\cline{2-13}
			&Yes&No&Open&Active&MPI&C/C++&OpenACC&Cuda&OpenCL&LIF&LZH&HH
			
			\cr
			\hline
			Brian2Cuda~\cite{goodman2008brian}
			&
			&{\bf{\checkmark}}  
			&{\bf{\checkmark}}
			&{\bf{\checkmark}}
			&{}
			& {\bf{\checkmark}}
			&{}
			& {\bf{\checkmark}}
			&{}
			& {\bf{\checkmark}}
			&{\bf{\checkmark}}
			& {\bf{\checkmark}}
			\cr\hline
			GeNN~\cite{stimberg2013brian}
			&{}
			&{\bf{\checkmark}} 
			&{\bf{\checkmark}}
			& {\bf{\checkmark}}
			&{}
			& {\bf{\checkmark}}
			&{}
			& {\bf{\checkmark}}
			&{}
			& {\bf{\checkmark}}
			&{\bf{\checkmark}}
			& {\bf{\checkmark}}
			\cr\hline
			Neuron~\cite{kumbhar2019coreneuron}
			&{\bf{\checkmark}} 
			& 
			&{\bf{\checkmark}}
			&{\bf{\checkmark}}
			&{\bf{\checkmark}}
			& {\bf{\checkmark}}
			&{\bf{\checkmark}}
			&{\bf{\checkmark}}
			&{}
			& {\bf{\checkmark}}
			&{\bf{\checkmark}}
			&{\bf{\checkmark}}
			\cr\hline
			HRLSim~\cite{6579754}
			&{\bf{\checkmark}}
			& 
			&
			&
			&{\bf{\checkmark}}
			& 
			&
			&  
			& 
			&{\bf{\checkmark}}
			& {\bf{\checkmark}}	
			&  
			\cr\hline
			{ NCS6 ~\cite{hoang2013novel}}	&{\bf{\checkmark}}
			&  
			&{\bf{\checkmark}}
			&
			&
			& {\bf{\checkmark}}
			&
			& 
			&{\bf{\checkmark}}
			&{\bf{\checkmark}}
			& {\bf{\checkmark}}
			&  
			\cr\hline
			
		\end{tabular}
		\begin{tablenotes}
			\footnotesize
			\item[*] There are three main categories of SNN models: Leaky Integrate and Fire (LIF), Izhikevich (IZH), and conductance-based models of which the Hodgkin-Huxley (HH).
		\end{tablenotes}
	\end{threeparttable}
\end{table*}

A variety of simulation applications have been developed to employ parallel technologies for better utilization of the multi-GPU architecture in supercomputers~\cite{chatzikonstantis2019multinode}.  
The simulator Brian2CUDA~\cite{goodman2008brian} generates C++/CUDA code to run on NVIDIA GPUs. GeNN~\cite{stimberg2013brian} is a code generator that can generate CUDA kernels and runtime code to be executed on multiple GPUs. Similar to GeNN, ANNarchy is a Python parallelizing interface used to create neural networks implementing neurons and synapses in the same manner as a brain. GeNN, Brian2CUDA, and ANNarchy are used to generate an entire C++ library optimized for parallel frameworks, such as OpenMP or CUDA, in a single GPU. However, simulations of large-scale brain models that contain over 1 billion neurons usually leverage a supercomputer as the computing platform to reduce simulation time and memory footprint. CoreNEURON~\cite{kumbhar2019coreneuron} is a compute engine for the NEURON simulator that is optimized for architecture with multiple CPUs and GPUs. Another high-performance simulator, HRLSim~\cite{6579754}, achieves a cognitive model in a supercomputer with a form and architecture similar to mammal brains. Message Passing Interface (MPI) is employed by HRLSim for both intra-node and inter-node communications. The neural cortical simulator (NCS6)~\cite{hoang2013novel} has been developed as a real-time neural simulator that may better utilize the heterogeneous clusters of CPUs and CUDA-capable GPUs in supercomputers. 
Table I gives a summary of the main features of the parallel technologies employed by the reviewed brain simulators in the multi-GPU architecture. CarlSim, CoreNeuron, HRLSim, and NCS6 are some of the representative brain simulators that support multi-GPU implementations.

Although encouraging progress has been made for brain simulations, the existing works focus on simulating mammal brains that have much simpler brain structures and fewer neurons compared to human brain. Although the latest work reported in~\cite{yamazaki2021human} simulates an entire human brain, all the simulated neurons are copies of a single neuron without considering physiology features of the real human brain. Communications do not form a performance bottleneck in the aforementioned simulation with a relatively small scale (for mammal brains) or a highly simplified human brain model. In order to reflect the communication models in real brains, which tend to have peer-to-peer data exchange between any pair of neurons, current brain simulators typically employ random assignment of neurons onto to GPUs and provide direct peer-to-peer connections between GPUs for the logical data flows between all neurons assigned on different GPUs. For example, the state-of-the-art brain simulators presented in~\cite{van2020exploring} and~\cite{yamazaki2021human} assign neurons to multi-GPUs randomly.

However, in a large-scale simulation for human brain data communications based on random neuron-GPU mapping and direct inter-GPU connections may cause network congestion inside a supercomputer thus introducing significant simulation latency. Therefore, more sophisticated schemes are needed for partitioning the simulated neurons to GPUs and controlling data transmissions between GPUs (which is referred to as routing in this paper). Low-latency communication design for brain simulations faces unique challenges -- the partitioning and routing methods need to minimize the number of connections between GPUs and balance the traffic across the connections on the one hand while maintaining the communication mode in the real brain as much as possible.     

There are a few algorithms that could be applied for partitioning neurons and grouping GPUs in brain simulations for low-latency communications. The minimum $k$-cut algorithm that finds the minimum-weight $k$-cut in a weighted graph may be applied to assign neurons to GPUs. However, analysis on the communication features of brain simulations in a multi-GPU environment indicates that minimizing the total traffic on the inter-GPU connections (which can be achieved by a $k$-cut algorithm) is not as effective as balancing traffic across these connections for latency reduction. Furthermore, the time complexity of the k-cut algorithm in a large-scale neural network with 10 billions of nodes make it not practical for real-time simulations~\cite{k-cut-approximation-2020}. Some meta-heuristic algorithms, including the genetic algorithm, simulated annealing algorithm, ant colony optimization, and swarm particle optimization, may be employed for addressing the partitioning problem. We evaluated these algorithms in our project and found that although a certain degree of traffic balance may be achieved the effectiveness of these algorithms for latency reduction is very limited in large-scale brain simulations, mainly due to the huge solution spaces that these kind of algorithms have to search through. The limitation of the existing methods motivated our development of new algorithms for neuron partitioning and traffic routing for achieving low-latency communications in large-scale brain simulations.

\section{Guidelines for Low-latency Communication Design in Brain Simulations} \label{Guidelines} 


The huge volume of floating-point operations for simulating a brain model makes brain simulation a computing-intensive application whose performance may be improved by allocating more computing resources, which is typically realized by utilizing more GPUs in a supercomputer. This implies that the neurons of the brain model must be dispersed across GPUs thus requiring data to be exchanged among GPUs. Although using more GPUs may acquire more computing capacity and larger memory space, distributing neurons across a large number of GPUs introduces overheads of inter-GPU communications. Therefore, communications may become a performance bottleneck in a supercomputer with a massive cluster of GPUs for brain simulations. 

Our analysis found that communications in brain simulations have the following characteristics.

\begin{enumerate}[1)]  
	\item The volume of data generated in brain simulations for information exchange among neurons is considerable, often in the magnitude of petabytes per second, which is very challenging to even the most advanced supercomputers.	
	
	\item The data transmissions in brain simulations usually require extremely low latency that might not be guaranteed by the conventional network control mechanisms. A main reason for latency in brain simulations is network congestion caused by the huge amount of traffic for inter-neuron communications.
	
	\item The communications among neurons for simulating a certain brain function are not evenly distributed. For example, some neurons mainly communicate with a few neurons while other neurons exchange information with a large number of neurons. The uneven distribution of inter-neuron traffic is a main factor that causes network congestion in brain simulations.
\end{enumerate}

Based on the above analysis, we suggest the following guidelines of communication design for brain simulations.
\begin{enumerate}[1)]  
	\item Network congestion is a main reason for the delay of multi-GPU communications in supercomputers. Considering the uneven traffic generated by bain simulations, the two key aspects of communication design -- partitioning for assigning neurons to GPUs and routing for steering data traffic among GPUs -- both should consider balancing inter-GPU traffic in order to avoid network congestion. 
	
	\item The communication design should attempt to cluster the neurons and GPUs in such a way that minimizes the traffic among GPUs for reducing communication latency. To achieve this objective, the partitioning algorithm should assign the neurons that have strong information exchange relationship onto the same GPU and the routing scheme should cluster the GPUs that have stronger communication demands among each other into the same group.  
	
	\item The number of connections among GPUs for inter-neuron data transmissions is also a key factor impacting simulation delay. A large number of inter-GPU connections cause significant overheads that may slow down or even halt the simulation process. Therefore, communication design for brain simulations should reduce the number of connections between each pair of GPUs as much as possible.
	
	\item Communication design for brain simulations has to enhance simulation performance by leveraging the current supercomputer architecture, which is difficult to be revised. Also, the design should strive to balance between adding control mechanisms (partitioning and routing) and maintaining the communication modes of brain models (peer-to-peer communications between neurons).      
\end{enumerate}

\section{A low-latency communication design} \label{Design}
 
Following the design guidelines given in the previous section, we propose in this section a low-latency communication design, which consists of a partitioning algorithm for assigning neurons to GPUs and a two-level routing method for controlling inter-GPU data transmissions. 

\subsection{Partitioning Algorithm}

The proposed partitioning scheme assigns neurons to GPUs in order to minimize the traffic between different GPUs. Since communications between neurons on the same GPU are much faster than that between neurons across GPUs, this algorithm reduces communication latency by assigning a set of neurons that have strong communication demand between each other on the same GPU. Furthermore, this algorithm makes the traffic between each pair of GPUs in the system as balanced as possible to avoid network congestion thus improving delay performance. 

\begin{algorithm}[!ht]
 	\caption{The partitioning algorithm}
 	\label{alg:A}
 	\begin{algorithmic}[1]
 		\REQUIRE~~\\ $P[M,M]$(the probability matrix of neurons connections), $W[M,1]$(the weight of neurons), $N$(the number of GPU in the system)
 		
 		\ENSURE~~\\ $PM$ (the neuron-GPU mapping)
 		\STATE Set parameters $T$(itermax) , $t$(current iteration) and Initialize all these parameters;
 		
 		\WHILE{$t$ $<=$ $T$ }
 		{  \STATE \bf	If}  $\sum_{i=1}^{N} w_{i} < avg \sum{W/N}$ 
 		{  \STATE \bf \hspace*{0.1in}	for}  $i = 1$ to $N$ {\bf do}
 		{\STATE \hspace*{0.3in} Greedy neurons assignment for every GPU by the probability matrix of neurons connections; }
 		
 		{\STATE \hspace*{0.3in} Select $x_{i,t}$ as  $w_{i} \in W_{i}$}
 		{\STATE \bf \hspace*{0.1in} end for}	
 		{\STATE  $t= t+1$}		
 		\ENDWHILE
 		{ \STATE Update the best optimal solution of the neuron-gpu mapping} 
 		{\STATE  \bf  Output $PM$}
 	\end{algorithmic}
 \end{algorithm}

Algorithm 1 gives the pseudocode of the proposed partitioning scheme. 
While maintaining the balance of the volume of each GPU, the algorithm uses a greedy strategy to assign neurons to every GPU to ensure high cohesion among neurons within a GPU and low coupling between neurons on different GPUs. The inputs for the algorithm include a matrix $P[M,M]$ and a vector $W[M,1]$ for a simulation model with $M$ neurons running on $N$ GPUs. The matrix $P[M,M]$ gives inter-neuron connection probability; i.e., each element $P[i,j]$ ($i, j \in [1, 2, \cdots, M]$) has a value in $[0, 1]$ representing the connection probability between the neurons $i$ and $j$. The $W[M,1]$ gives the weight of each neuron that describes the data volume generated by the neuron for communications with all other neurons. $x_{i,t}$ represents the output of Algorithm 1 in each iteration, which is the neuron that needs to be placed in each GPU. The goal of the algorithm may be formulated as $min \ \Sigma_{i,j} (P[i,j]*W[i]*W[j])$ for all $i$ and $j$ that are not assigned to the same GPU; that is, minimizing the total amount of traffic across different GPUs for inter-neuron communications. During each iteration, the algorithm first balances the total weight of all the neurons assigned to each GPU and then selects the best optimal solution of the neuron-GPU mapping using a greedy assignment for each GPU based on the probability matrix. Finally, the algorithm updates the solution and outputs a table for neuron-GPU mapping.

\subsection{Two-Level Routing Method}

 \begin{algorithm}[!t]
	\caption{The two-level routing method}
	\label{alg:b}
	\begin{algorithmic}[1]
		\REQUIRE~~\\  $PG[N,N]$ (the probability matrix of GPUs connections),  $WG[N,1]$ (the sum weight of neurons in each GPU)
		
		\ENSURE~~\\  $TB$ (the routing table)
		\STATE Set parameters $T$(itermax), $t$(current iteration), $G$(cluster groups) and Initialize all these parameters;
		\STATE Set $PG[N,N]$ and $WG[N,1]$ according to the partitioning algorithm;
		
		\WHILE{$t$ $<=$ $T$ }
		{  \STATE \bf	while}  $\sum_{i=1}^{N} w_{i} < avg \sum{WG/N}$ 
		{  \STATE \bf \hspace*{0.1in}	for}  $i = 1$ to $N$ {\bf do}
		{\STATE \hspace*{0.3in} Greedy GPUs assignment for every group $G_{i}$ by the probability matrix $PG[N,N]$; }
		{\STATE \hspace*{0.3in} GPU connections in $G_{i}$ are the level-1 routing; }
		{\STATE \hspace*{0.3in} Select GPUs to connect other groups $G_{j}$ as the level-2 routing; }
		
		{\STATE \bf \hspace*{0.1in} end for}	
		{\STATE \hspace*{0.1in} $t= t+1$}
		{\STATE \bf \hspace*{0.05in} end while}		
		\ENDWHILE
		{ \STATE Update the best optimal solution of the routing method} 
		{\STATE  \bf  Output $TB$}
	\end{algorithmic}
\end{algorithm}

Although the partitioning algorithm reduces traffic between GPUs, the large number of inter-neuron connections across GPUs with uneven traffic distribution is still a challenge that must be addressed for achieving low-latency communications. In a large scale brain simulation, thousands of GPUs attempt to communicate with each other simultaneously thus may cause network congestion that degrades delay performance. 
For example as illustrated in Figure 2(a), at a time instant GPUs No. 1 to No. 8 all need to communicate with GPU No. 12. If the total traffic amount is greater than the link capacity to GPU No. 12, then the inevitable congestion on the link will introduce extra latency to the simulation functions running on these GPUs. Please note that the connections that we are referring to here are for logical data flows (shown as red lines in Figure 2) while the physical data flows (shown as green lines in Figure 2) follow the network structure, e.g., an Infiniband network, in the supercomputer. 

For addressing this challenge to low-latency communications, we propose a control scheme with a two-level routing structure for reducing the number of connections between GPUs while balancing the traffic distribution across the connections. This scheme clusters GPUs in the system into groups with an objective of minimizing the amount of traffic between GPUs in different groups. GPUs in the same group communicate with each other through direct connections. When a GPU wants to communicate with another GPU in a different group, it first transmits data to a bridge GPU in the same group and then the bridge GPU forwards data to the destined GPU. As illustrated in Figure 2(b), the two-level routing scheme with GPU clusters reduces the number of direct connections between GPUs. Please note that although the physical data flow structure stays the same (as shown by the green lines in Figure 2 (a) and (b)) the amounts of traffic on these flows are reduced by the two-level routing scheme. It is worth noting that the proposed two-level routing is an application layer function for steering the traffic of logical data flows, which is different from the network routing in the underlying supercomputer architecture that controls the forwarding of physical data flows through switches. A brain simulator runs as an application on top of the supercomputer architecture; therefore, the communication control mechanism in the simulator leverages the supercomputer infrastructure including its routing protocol.

While maintaining the balance of the
traffic of each group and the number of direct connections between GPUs, the proposed method also uses a greedy strategy to assign GPUs to every groups to ensure high cohesion among neurons within a group and low coupling between neurons on different groups. The reason is GPUs in the same group communicate with each other through direct connections, and GPUs in different groups need to identify the corresponding GPU node in their group to forward data traffic. Moreover, a group of GPUs on the same switch in the physical structure, which speed of the information interaction is faster. In addition, because each connection requires a thread to be started, the time taken to start the thread in the whole system can be reduced by reducing the number of connections.
Algorithm 2 describes the proposed routing method. Inputs to the algorithm include the matrix $PG[N,N]$ and vector $WG[N,1]$, where $N$ denotes the number of GPUs in the system. Each element $PG[i,j]$ of the matrix $PG[N,N]$ represents the connection probability between GPUs $i$ and $j$ ($i, j \in [1, N]$). The $WG[N,1]$ gives the total weight of all the neurons assigned on each GPU, which describes the data volume generated by the GPU. The goal of the algorithm is to minimize the number of inter-GPU connections and reduce the traffic across GPU groups, which can be formulated as $min \ Conn[i,j]$, where $i,j \in [1,N]$ and $Conn[i,j]$ denotes the number of connections between GPUs $i$ and $j$, and $min \ (\Sigma_{i,j}PG[i,j]*WG[i]*WG[j])$ for all pairs of GPUs $i$ and $j$ that are not in the same group. 
In each iteration of the loop (lines 3-12) the algorithm updates the best optimal solution of the routing by adapting the greedy assignment for every GPU group $G_{i}$ to cluster the most connected GPUs in the same group. GPUs in the same group are connected to each other via the level-1 routing (i.e. direct peer-to-peer connections). If a GPU wants to connect to a GPU in another group and itself is not a bridge node on behalf of its own group, then it needs to find another GPU as the bridge node for data forwarding. In the end, the algorithm obtains a routing table that defines the connection relationship for each GPU as the output.

\section{Experiment Results} \label{Experiment}

In this section, we evaluate the performance of the proposed partitioning and the routing methods for brain simulations. In the experiments, we used 2,000 GPUs on the supercomputer Kunlun to simulate a brain model with 10 billion neurons. The model was created according to the biological structure of a real human brain scanned using medical instruments. The simulation involves memory transfer and communications between GPUs via PCIE and InfiniBand. We used OpenMPI, which combines two parallel programming methods MPI and CUDA to achieve the highest data transmission efficiency. For performance evaluation, we compared the data traffic and number of connections between GPUs with and without the proposed design. We also evaluated the simulation time to verify the effectiveness of the proposed design for low-latency communication. 
  \begin{table*}[!t]
 	\centering
 	\fontsize{8}{16}\selectfont
 	\caption{Latency performance comparison of different simulation scales}
 	\label{tab:performance_comparison}
 	\begin{tabular}{p{1.2cm}c|c|c|c|c|c|c|c|c}

 		\hline
 		
 		\multicolumn{9}{c}{\bf{\qquad\qquad\qquad\qquad\qquad\qquad\qquad\qquad\qquad\qquad\qquad\qquad\qquad\qquad The channel noises in brain simulations }}\cr\cline{2-10}
 		
 		\multirow{1}{*}{\bf{Scales}}&
 		\multirow{1}{*}{\bf{Neurons}}&
 		\multirow{1}{*}{\bf{Partition }}&
 		\multirow{1}{*}{\bf{Routing }}&
 		\multicolumn{1}{c}{\bf{0.1}}&
 		\multicolumn{1}{c}{\bf{0.2}}&
 		\multicolumn{1}{c}{\bf{0.3}}&
 		\multicolumn{1}{c}{\bf{0.4}}&
 		\multicolumn{1}{c}{\bf{0.5}}&
 		\multicolumn{1}{c}{\bf{0.6}}\cr\cline{2-10}

 		\hline
 			{\bf {2000GPUs}}&{10 billion}&{Random}&{P2P}&{4h30min+}&{4h30min+}&{4h30min+}&{4h30min+}&{4h30min+}&{4h30min+}\cr\hline 
 			 
 			{\bf {2000GPUs}}&{10 billion}&{GA}&{GA}&{4h20min+}&{4h20min+}&{4h20min+}&{4h20min+}&{4h20min+}&{4h20min+}\cr\hline

 		{\bf {2000GPUs}}&{10 billion}&{proposed}&{proposed}&{0.179s}&{0.27s}&{0.319s}&{0.339s}&{0.357s}&{0.367s}\cr\hline

 		{\bf{4000GPUs}}&{20 billion}&{proposed}&{proposed}&{0.323s}&{0.377s}&{0.413s}&{0.499s}&{0.467s}&{0.491s}\cr\hline
 	\end{tabular}
 \end{table*}
 
Comparison of inter-GPU traffic generated by simulations with different neuron partitioning methods is presented in Figure 3(a). In this figure, the red, blue, and green lines give the amounts of traffic generated by each of the 2000 GPUs in the simulation system when the system uses random mapping, a genetic algorithm, and the proposed algorithm respectively for partitioning. It can be seen from the red line that random partition causes significant fluctuation in the amounts of traffic from different GPUs, which may easily cause congestion on some network links in the supercomputer. The blue line indicates partition using a genetic algorithm may obtain a certain degree of balance for inter-GPU traffic but not sufficient yet to avoid network congestion for reducing simulation latency (as shown by the results reported in Table II). The green line shows that the proposed partitioning algorithm may greatly reduce the difference in traffic volumes across GPUs. The peak value of the green line is 31.2\% lower than that of the red line and 13.4\% lower than  that of the blue line. Therefore, the proposed algorithm is more effective in balancing traffic and avoiding congestion compared to current methods. We also evaluated other heuristic algorithms, including simulated annealing, ant colony optimization, and swarm particle optimization, for neuron partitioning in our experiments and found they achieved similar performance as the genetic algorithm, which is used in Figure 3(a) as a representative.


Figure 3(b) shows the effectiveness of the proposed two-level routing method for controlling the amount traffic for inter-GPU communications. In this figure, the red line gives the amount of traffic generated from each of the 2000 GPUs when the two-level routing method is not employed. The blue line and green line in this figure respectively give the amounts of level-2 traffic when the two-level routing scheme is employed and the GPUs are clustered using a genetic algorithm and the proposed algorithm. Level-2 traffic, i.e., traffic destined to GPUs in other groups, is used in this figure because it has a larger impact on communication delay compared to level-1 traffic within each GPU group.
This figure shows that the proposed algorithm may achieve better balanced level-2 traffic across different GPUs than what a genetic algorithm does (The peak value of the blue line is $39.2\%$ lower than that of the green line). Moreover, the peak value of the green line is $51.1\%$ lower than that of the red line, it shows the effectiveness of the two-level routing method.
  Because simulation latency is determined by the communication between the pair of GPUs that has the largest traffic volume, balancing inter-GPU traffic may greatly improve delay performance for brain simulations. 

Our analysis indicates that a key aspect of achieving low-latency communication in brain simulations is to reduce the total number of connections between GPUs. In order to evaluate the performance of the proposed method in this aspect, we present the histograms of frequency distribution for the total number of inter-GPU connections generated in the simulation system using direct peer-to-peer routing and the proposed two-level routing respectively in sub-figures (a) and (b) of Figure 4. This figure shows that the two-level routing method proposed in this paper is able to reduce the number of connections between each GPU and all its destined GPUs. Our experiments found that the average number of connections departing from each GPU is reduced from 1,552 to 88 by using the proposed routing method. 

In order to evaluate the delay performance and scalability of the proposed communication design, we tested the simulation time for the brain model with various levels of computational complexity and system scales. The obtained results are reported in Table II. The channel noise in the simulation is a parameter that reflects the complexity of simulating brain functions. As this parameter increases from 0.1 to 0.6, the computational complexity for brain simulation increases significantly. 

We first run the simulation on 2000 GPUs using random neuron-to-GPU partition and peer-to-peer routing between GPUs. The obtained latency results are listed in the first row of Table II, which give very long latency (more than four and half hours) that makes the simulation infeasible in practice. We tested the simulation time when we explored heuristic algorithms, such as the genetic algorithm (GA), simulated annealing algorithm, ant colony optimization, and swarm particle optimization, for partitioning neurons and clustering GPUs and found they all achieved the same level of performance. The results obtained using GA are presented in the second row of Table II, which indicates that basically no improvement in simulation latency is achieved by applying such algorithms in communication control.

Our experiments show that the simulation time is reduced significantly when the proposed partitioning and routing algorithms are employed in the simulation system. The obtained results for a model with 10 billion neurons running on 2000 GPUs are given in the third row of Table II. The data show that the simulation latency achieved for all complexity levels are less than one second, which satisfy the requirement of real-time inter-neuron communications in human brain. We also run the simulations for 20 billion neurons using 4000 GPUs on the same supercomputer and the measured results of simulation time are shown in the forth row of Table II. This row of data shows that the proposed partitioning and routing schemes are able to ensure low-latency communications even in a super-large simulation scale. The last two rows in Table II also verify that our communication design is scalable with respect to the brain model (from 10 billion neurons to 20 billion neurons), the simulation system (from 2000 GPUs to 4000 GPUs), and the simulation complexity (channel noise parameter from 0.1 to 0.6). We also noticed that the latency increment with respect to channel noise is smaller in the model with 20 billion neurons than that in the model with 10 billion neurons. That is mainly caused by the biological features of the simulated brain -- the synapses among the 20 billion neurons are sparser compared to those among the 10 billion neurons. Therefore, when the model with 20 billion neurons is simulated on 4000 GPUs it generates less intensive traffic for inter-GPU communications than what 10 billion neurons generate on 2000 GPUs, which leads to less latency increment as the channel noise increases.

\section{Challenges and future directions}  \label{Directions}

In this section, we shed light on several challenges to communication design in large-scale simulations of low-latency applications such as brain modeling.

\subsection{Modifications to MPI Library }

Communications in low-latency applications such as brain simulations require appropriate assignment of (virtual) destination LIDs for MPI~\cite{kang2020improving}, which cannot be satsified by existing MPI libraries used in supercomputers. Although we used OpenMPI to support IB’s multi-LID addressing, this is achieved in the form of a static topology table, which cause extra communication latency when it is used for mapping datasets to GPU devices and managing topology. Additionally, existing communication scheduling algorithms of MPI are relatively simple, which directly determine the calls of communication functions such as MPI-Send and MPI-Recive.

The communication kernel for redistributing requests in MPI development provides a many-to-many communication mode for applications with numerous discontinuous requests, which plays a crucial role in the overall system performance. Thus, we can further improve the communication performance of asynchronous MPI implementations, and adopt the communication throttling method to reduce the number of communication contentions to reduce inter-GPU communication delay.

\subsection{Parallelization of algorithms}

Parallel algorithm design plays a crucial role in enhancing performance of simulation applications in supercomputers with multi-GPUs. These algorithms include analog computing functions (such as SNN) and the algorithms studied in this paper~\cite{9748023}. An effective algorithm parallel design divides the models and algorithms into different layers and distributes them to different hardware, while achieving load balance and reducing communication overheads. Both parallelism and scalability should be considered during the algorithm design process. Different designs will directly affect the algorithm running speed on a supercomputer. A preliminary test on our algorithm indicates that a parallel design may speed up the algorithm by more than 10 times.

\subsection{Larger scale simulations of human brain}

Simulation of human brain is a large-scale research project. Although we have designed a low-latency communication framework in the simulation of 10 billion neurons spread across 2,000 GPUs, the simulation realizes part of human brain functions. The secrets of human brain are far beyond our present capabilities. The human brain has 86 billion neurons and the simulation complexity increases exponentially with the number of simulated neurons. We estimate that tens of thousands of GPUs are needed to simulate the entire human brain. An even bigger challenge is that the connections among neurons in the brain do not form a regular network. If a matrix were used to represent the inter-neuron connections in the brain, the matrix would be extremely sparse, which brings in unique challenges to communication design in the simulation system for effectively simulate the rapid information exchange among neurons during the thinking process of human brain. 

\section{Conclusion}

In this paper, we investigated the problem of low-latency communications for supporting simulations of large-scale brain models. We first summarized a list of design guidelines for achieving low-latency communications based on our analysis on the communication characteristics in brain simulations. Then we proposed a communication design, including a partitioning algorithm for assigning neurons to GPUs and a two-level routing method for controlling data transmissions among the GPUs, for achieving low-late communications required by brain simulations. Experiment results reported in this paper show that the proposed partitioning algorithm and routing method together may significantly improve the delay performance of inter-GPU communications in brain simulations. We also discussed open issues and identify some research directions in design of a low-latency communication framework in supercomputers.
In our future work, we will continue to conduct large-scale human brain simulations to solve the problem of algorithm parallelization and MPI modification to further enhance the performance of brain simulations.

\bibliographystyle{IEEEtran}
\bibliography{references}

\end{document}